\documentclass[12pt]{elsart}

\usepackage{graphicx}
\usepackage{subfigure}
\usepackage{amsmath}
\usepackage{cite}
\usepackage{endfloat}
\usepackage{tikz}

\newcommand*\circled[1]{\tikz[baseline=(char.base)]{\node[shape=circle,draw,inner sep=2pt] (char) {#1};}}
\DeclareMathOperator*{\argmin}{arg\,min}

\journal{Physics Letters A}

\begin{document}

\begin{frontmatter}

\title{Ground states of the frustrated Blume-Emery-Griffiths model in a field}
\author{M. \v{Z}ukovi\v{c}}
\ead{milan.zukovic@upjs.sk}
\address{Department of Theoretical Physics and Astrophysics, Faculty of Science,\\ 
P. J. \v{S}af\'arik University, Park Angelinum 9, 041 54 Ko\v{s}ice, Slovakia}

\begin{abstract}
Ground-state properties of the Blume-Emery-Griffiths model with antiferromagnetic nearest-neighbor interactions on a triangular lattice are investigated in the presence of an external magnetic field. In particular, we explore the model's parameter space and identify regions with different degenerate ground states that may give rise to different magnetic phases also at finite temperatures. We demonstrate the presence of such phases by Monte Carlo simulations of magnetization processes for selected values of parameters. 
\end{abstract}


\end{frontmatter}

\section{Introduction}
\hspace*{5mm} A frustrated triangular lattice Ising antiferromagnet (TLIA) with spin 1/2 is long known to display no long-range ordering down to zero temperature~\cite{wann}, albeit the ground state is critical with the power-law decaying spin-correlation function~\cite{step}. Nevertheless, a long-range order can occur in the ground state if the spin is larger than some critical value, estimated as 11/2~\cite{naga,yama,lipo}. Generally, the lack of order in frustrated spin systems is due to large ground-state degeneracy. However, this can be lifted by various perturbations, such as an external magnetic field~\cite{metcalf,schick,netz,zuko} or selective dilution~\cite{kaya,zuko1}, which can result in long-range ordering even in TLIA with spin 1/2. In the Ising models with spin larger than 1/2, a single-ion anisotropy and higher-order (e.g., biquadratic) exchange interactions may play a crucial role in their critical properties (see, e.g.,~\cite{cape,blum,beg}). The model that incorporates the above mentioned effects is known as Blume-Emery-Griffiths (BEG) model~\cite{beg} and has a long history of investigation~\cite{beg,berk1,chen,netz1,wang,wang1,host,tuck,grig,akhe}. In the case of the BEG model with antiferromagnetic interactions, it is interesting to study its behavior in an external magnetic field. A recent study by using exact recursion relations on the Bethe lattice~\cite{erdi} produced some interesting results, such as the reentrant phenomenon with the consecutive phase transitions from disorder to order and back to disorder as the field was increased. However, as already observed for example in the simple TLIA model with spin 1/2, a frustrated antiferromagnet in the presence of an external field behaves quite differently than its nonfrustrated counterpart~\cite{metcalf,schick,netz,zuko}. A frustrated Blume-Capel model on a triangular lattice, which is special case of our BEG model with zero biquadratic exchange interactions and zero magnetic field, has been investigated by position-space renormalization group methods~\cite{maha} and has been shown to display finite-temperature long-range ordering within a certain range of the single-ion anisotropy strength, accompanied with a multicritical behavior. In order to understand finite-temperature behavior of the frustrated spin systems, which is often unexpected and intricate, it is important to understand their ground-state properties.\\
\hspace*{5mm} In the present Letter we consider the geometrically frustrated antiferromagnetic BEG model on a triangular lattice in the presence an external magnetic field. As a result of the frustration, the ground state of such a system is highly degenerate. On the other hand, the model features a number of different perturbations that can lift this degeneracy in different ways and thus we can expect much richer variety of magnetic structures in a broad parameter space than for a nonfrustrated case.

\section{Model and methods} 
\hspace*{5mm} We consider the spin-1 Ising model on a triangular lattice described by the Hamiltonian 
\begin{equation}
\label{Hamiltonian}
\mathcal H=-J_1\sum_{\langle i,j \rangle}S_{i}S_{j}-J_2\sum_{\langle i,j \rangle}S_{i}^2S_{j}^2-D\sum_{i}S_{i}^2-h\sum_{i}S_{i},
\end{equation}
where $S_{i}=\pm1,0$ is a spin on the $i$th lattice site, $\langle i,j \rangle$ denotes the sum over nearest neighbors, $J_1<0$ is an antiferromagnetic bilinear exchange interaction parameter, $J_2$ is a biquadratic exchange interaction parameter, $D$ is a single-ion anisotropy parameter, and $h$ is an external magnetic field. \\
\hspace*{5mm} Considering the presence of only nearest-neighbor interactions, it is sufficient to focus on an elementary triangular plaquette formed by the neighboring spins. Then, a reduced zero-temperature energy per spin can be expressed as 
\begin{equation}
\label{Energy}
e/|J_1|=\sum_{\langle k,l \rangle}S_{k}S_{l}-J_2/|J_1|\sum_{\langle k,l \rangle}S_{k}^2S_{l}^2-D/(3|J_1|)\sum_{k}S_{k}^2-h/(3|J_1|)\sum_{k}S_{k},
\end{equation}
where the summation $\langle k,l \rangle$ runs over the nearest neighbors $S_{k}$ and $S_{l}$ ($k,l=1,2,3$) on the plaquette.  Ground-state (GS) spin configurations ${\bf S_{\mathrm{GS}}}=\{S_1,S_2,S_3\}$ can be found as configurations\footnote{There are in total 27 configurations to be considered.} that minimize the energy functional $f({\bf S,P}) \equiv e/|J_1|$ with the parameters values ${\bf P} = (J_2/|J_1|,D/|J_1|,h/|J_1|)$, i.e.:
\begin{eqnarray}
\label{GS} {\bf S_{\mathrm{GS}}} = \argmin_{{\bf S}} \, f({\bf S, P}).
\end{eqnarray}
Apparently, for some parameters values there will be no single optimal solution but rather a multitude of degenerate states with the same energy. Then the GS energy per spin can be obtained as a function of the parameters $\bf P$: $f_{\mathrm{GS}}({\bf P})=f({\bf S_{\mathrm{GS}},P})$.\\
\hspace*{5mm} In order to demonstrate the presence and character of different phases resulting from the GS configurations found from Eq.~(\ref{GS}), we further employ Monte Carlo (MC) method and study magnetization processes in different regions of the parameter space. We perform MC simulations on a spin system of a moderate linear size of $L=24$, employing the Metropolis dynamics and applying the periodic boundary conditions. For thermal averaging we consider $N=5 \times 10^4$ MCS (Monte Carlo sweeps or steps per spin) after discarding another $10^4$ MCS for thermalization. For configurational averaging and calculating error bars we carry out five independent runs. The simulations are performed at a fixed reduced temperature close to the ground state ($t \equiv k_BT/|J_1|=0.1$) and a varying external field $h/|J_1|$. The simulations start from zero field, using a random initial configuration, and then the field is gradually increased with the step $\Delta h/|J_1|=0.1$ and the simulation starts from the final configuration obtained at the previous field value. The triangular lattice is considered to be consisting of three interpenetrating sublattices A, B and C, and we calculate the respective sublattice magnetizations per site
\begin{equation}
\label{sub_mag}
(m_{\mathrm{A}},m_{\mathrm{B}},m_{\mathrm{C}}) = 3\Big(\Big\langle\sum_{i \in \mathrm{A}}S_{i}\Big\rangle, \Big\langle\sum_{j \in \mathrm{B}}S_{j}\Big\rangle, \Big\langle\sum_{k \in \mathrm{C}}S_{k}\Big\rangle\Big)/L^2
\end{equation}
and the total magnetization per site
\begin{equation}
\label{mag}
m = \Big\langle\sum_{n=1}^{L^2}S_{n}\Big\rangle/L^2,
\end{equation}
where $\langle\cdots\rangle$ denotes thermal averages.

\section{Results}
\begin{figure}[t]
\centering
    \subfigure[$J_2/|J_1|=-1.25$]{\includegraphics[scale=0.47]{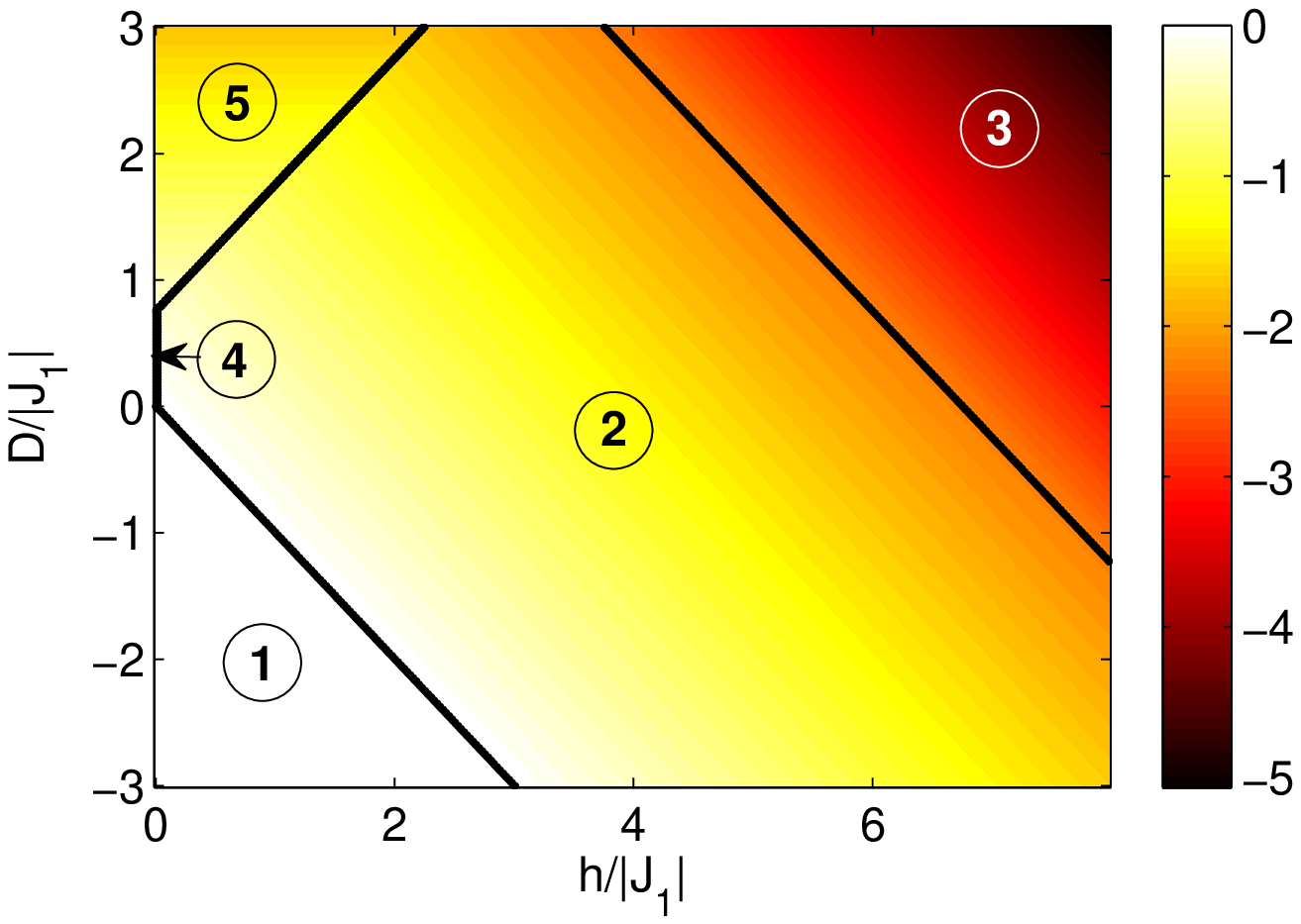}\label{fig:gse_j2_-1_25}}
		\subfigure[$J_2/|J_1|=-1.00$]{\includegraphics[scale=0.47]{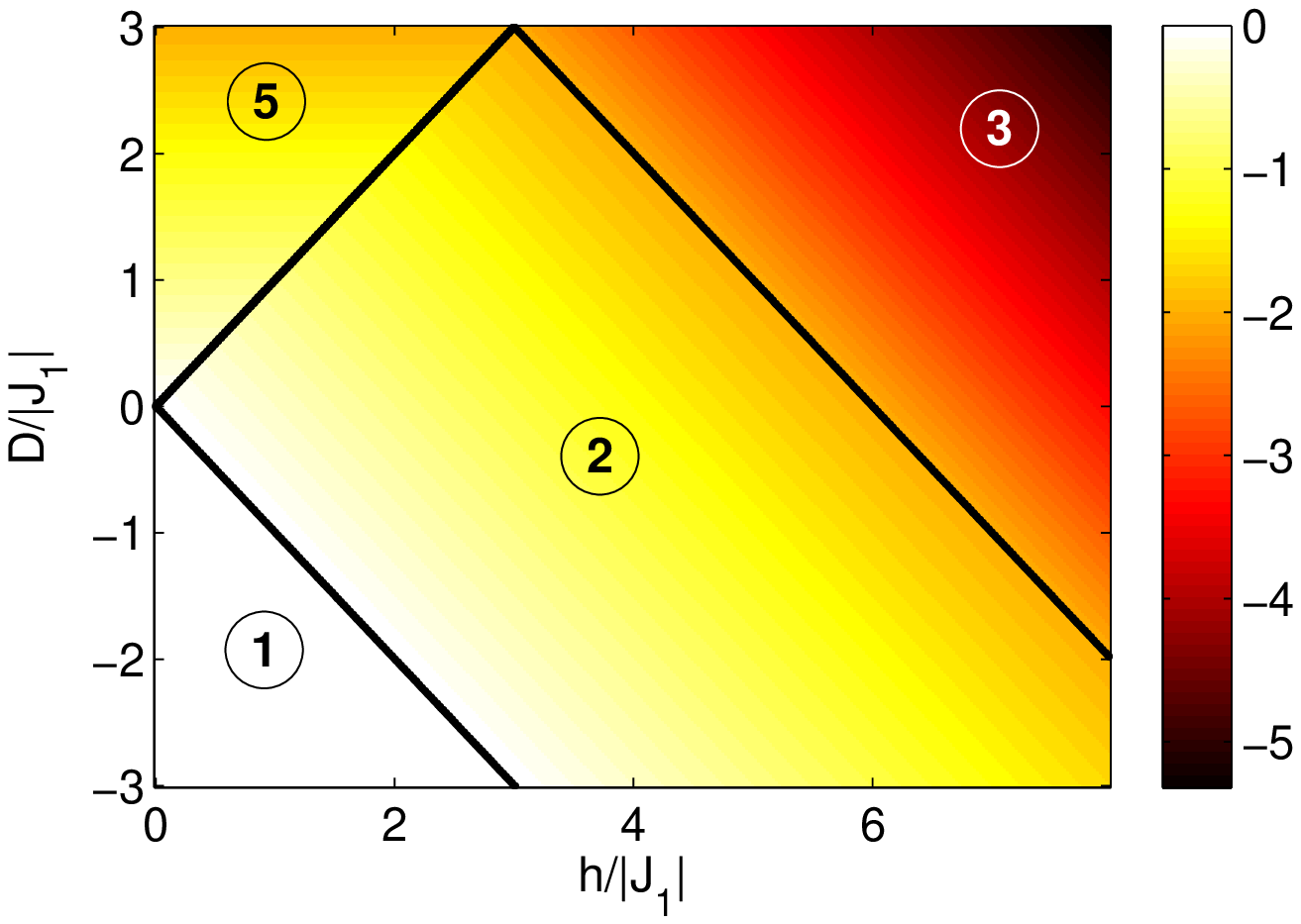}\label{fig:gse_j2_-1_00}}
		\subfigure[$J_2/|J_1|=-0.75$]{\includegraphics[scale=0.47]{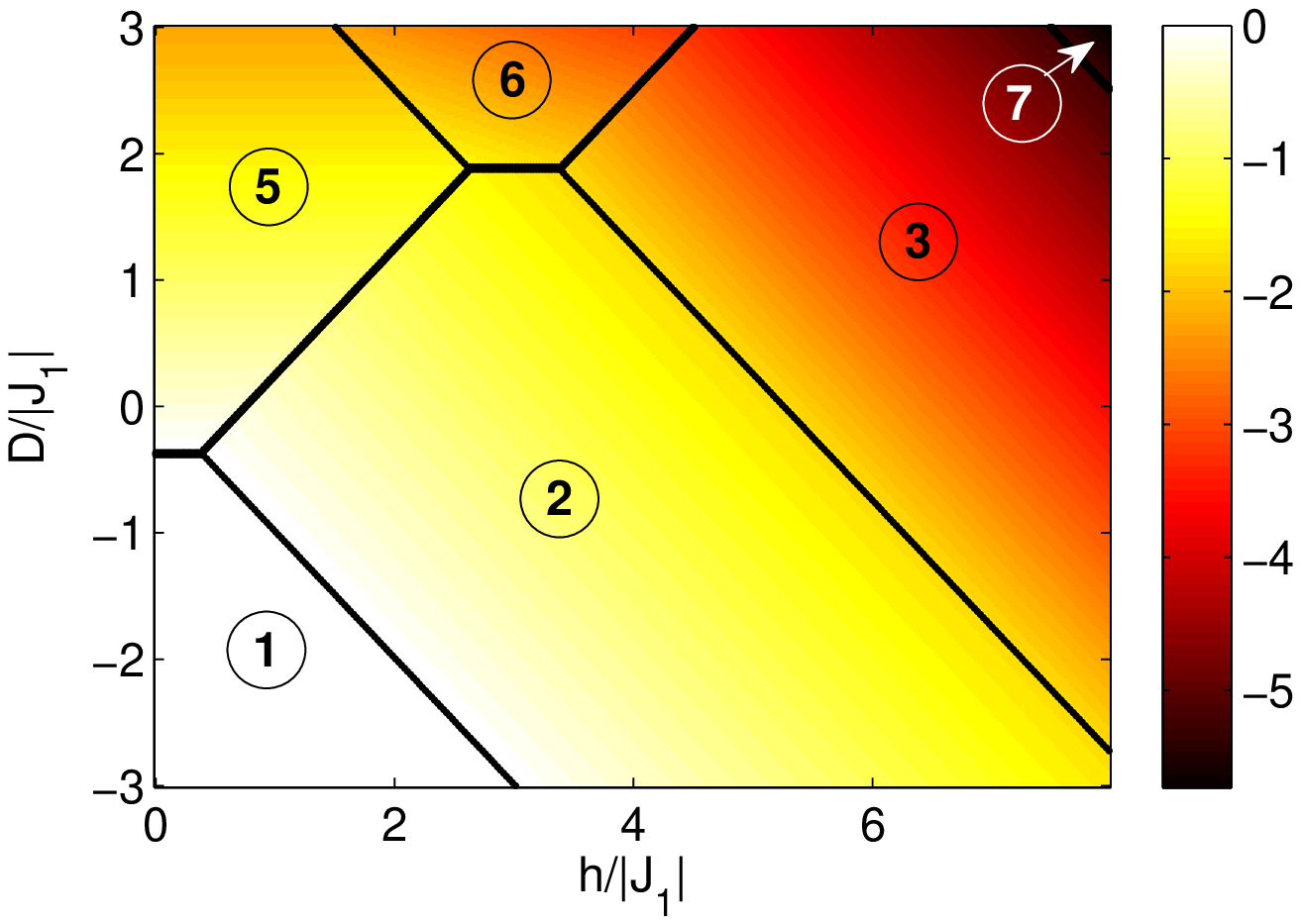}\label{fig:gse_j2_-0_75}}
		\subfigure[$J_2/|J_1|=-0.50$]{\includegraphics[scale=0.47]{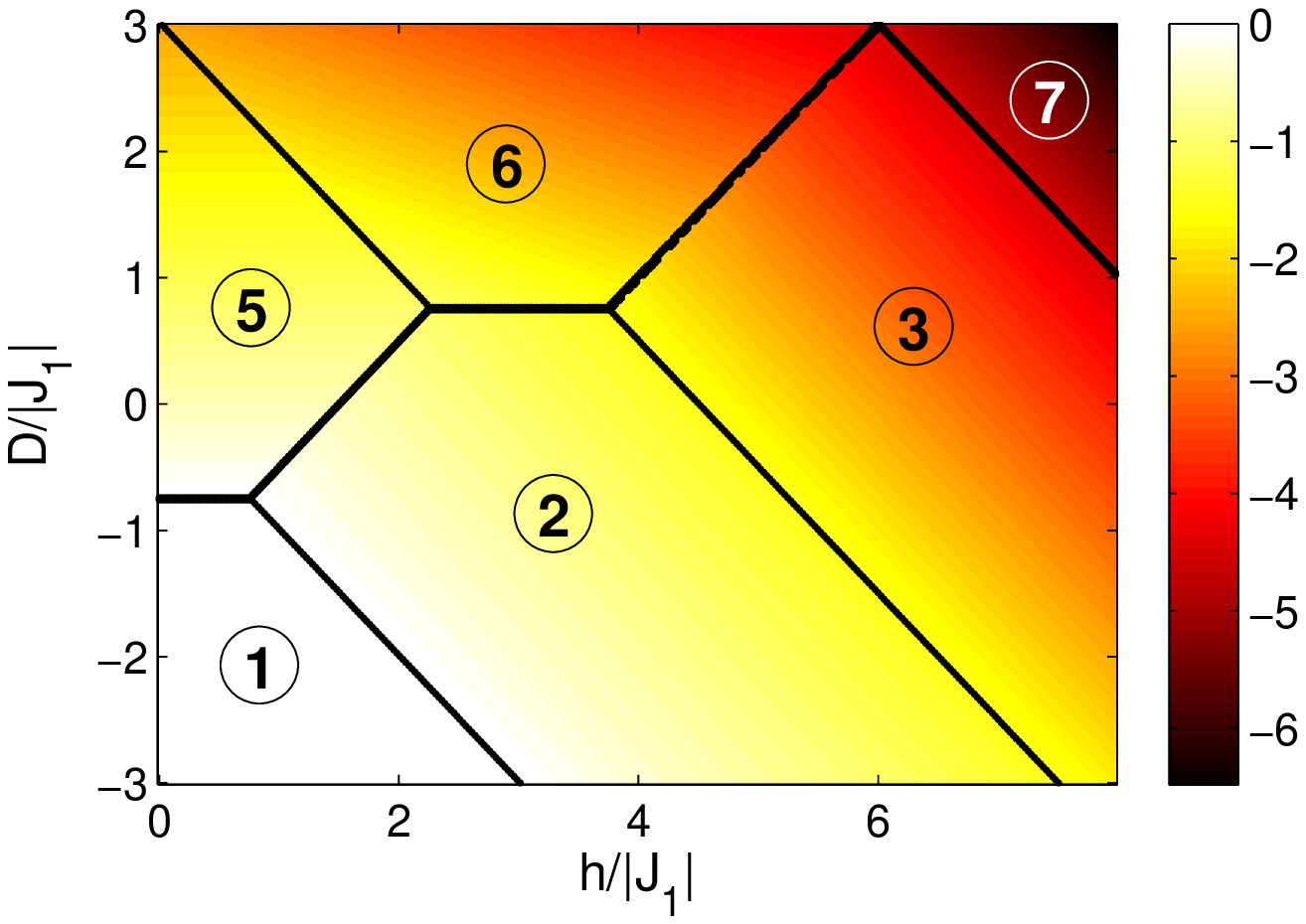}\label{fig:gse_j2_-0_50}}
		\subfigure[$J_2/|J_1|=-0.25$]{\includegraphics[scale=0.47]{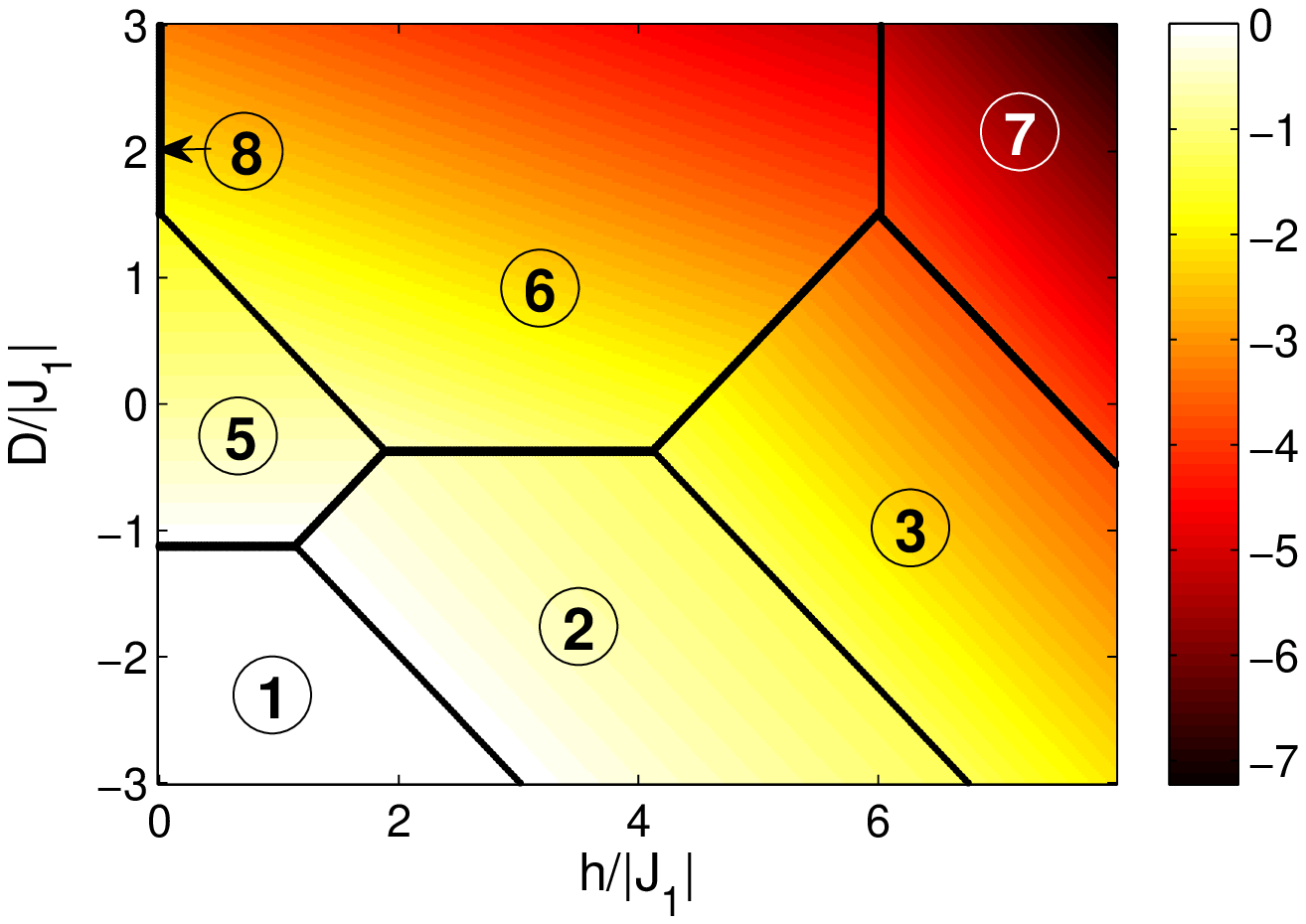}\label{fig:gse_j2_-0_25}}
		\subfigure[$J_2/|J_1|=0.00$]{\includegraphics[scale=0.47]{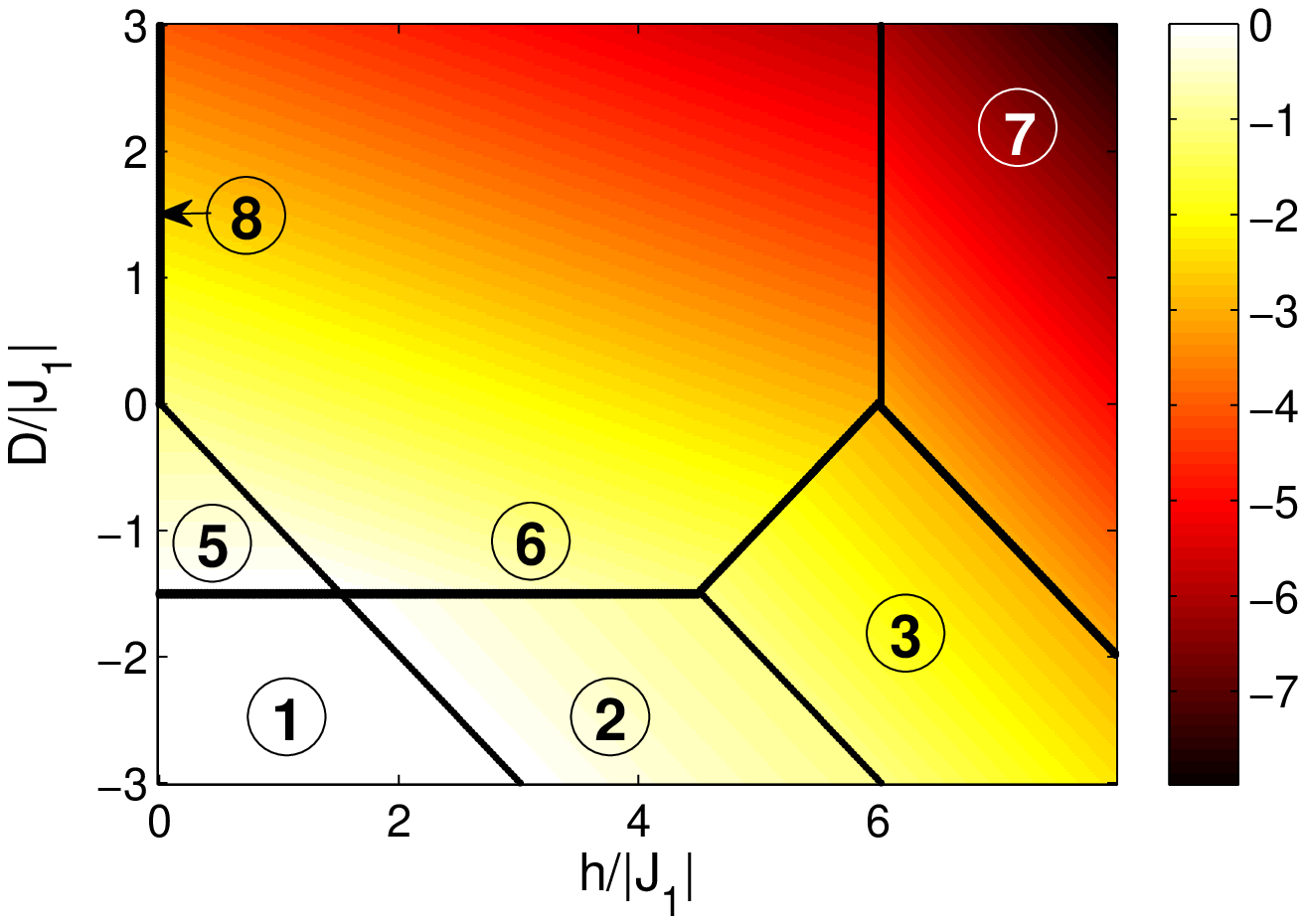}\label{fig:gse_j2_0_00}}
		\subfigure[$J_2/|J_1|=0.25$]{\includegraphics[scale=0.47]{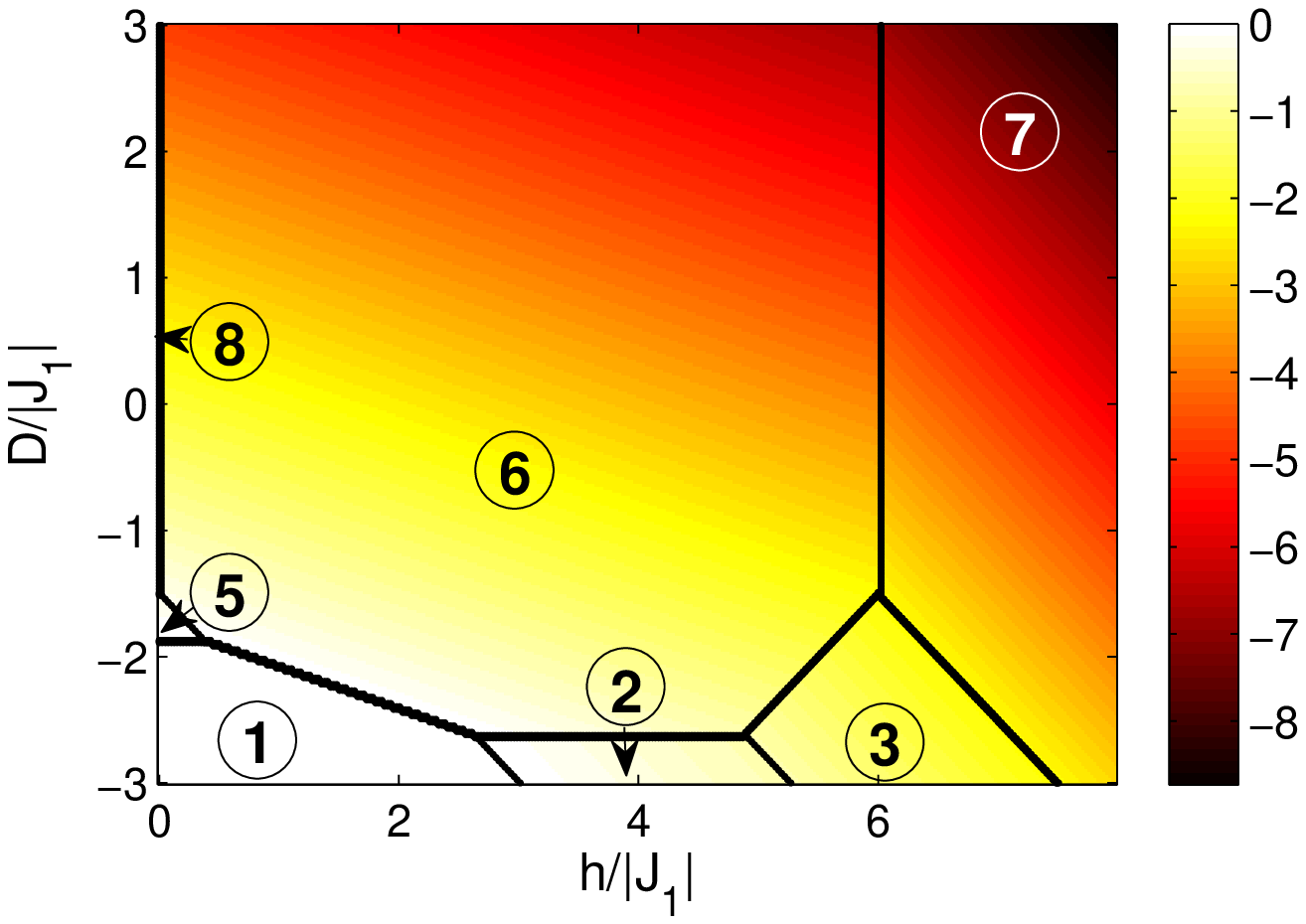}\label{fig:gse_j2_0_25}}
		\subfigure[$J_2/|J_1|=0.50$]{\includegraphics[scale=0.47]{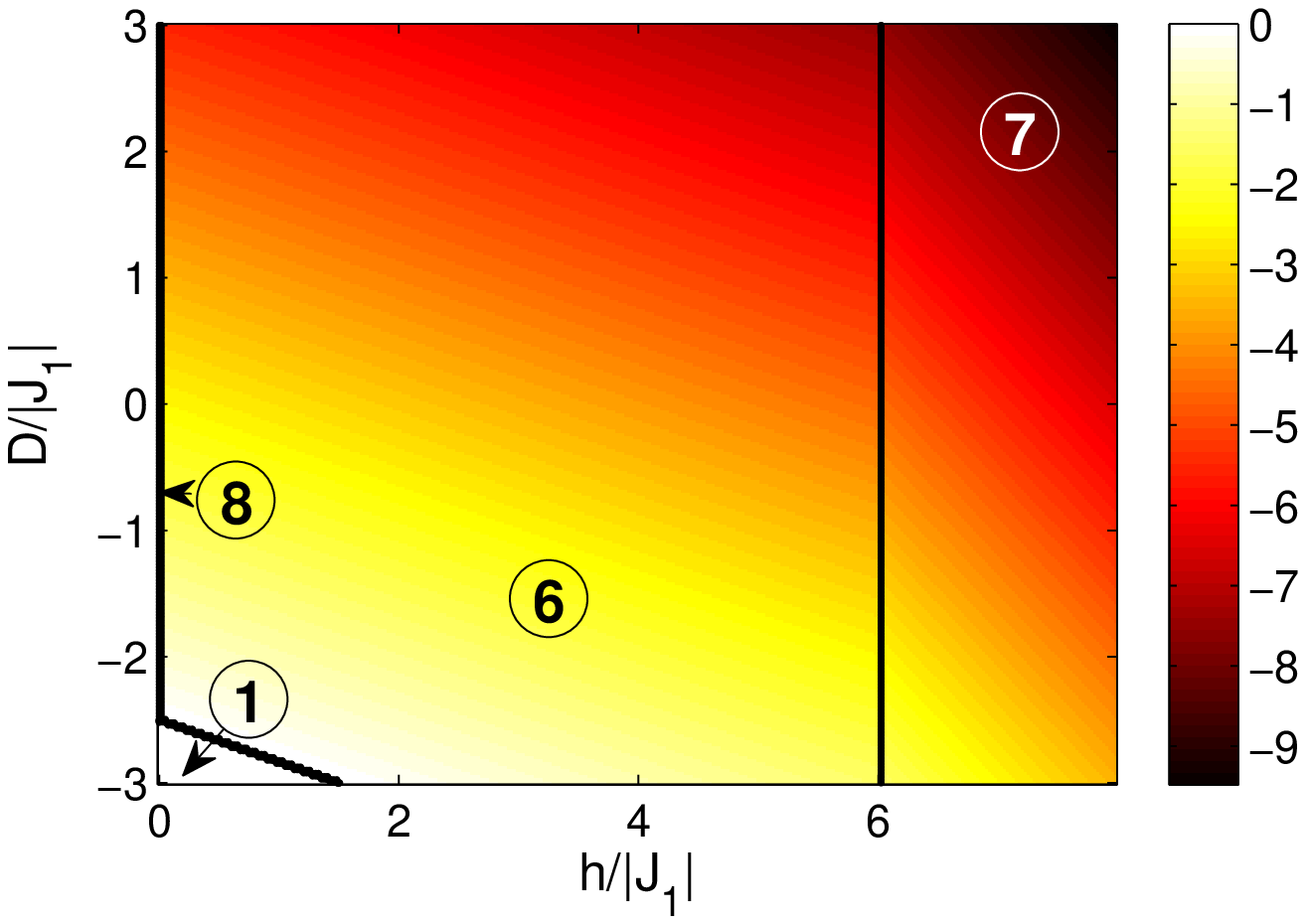}\label{fig:gse_j2_0_50}}
\caption{(Color online) Surfaces of the ground-state energy $f_{\mathrm{GS}}$ in the $D/|J_1|-h/|J_1|$ parameter plane for various values of the biquadratic exchange parameter $J_2/|J_1|.$ The lines mark the borders between different phases, presented in Table~\ref{tab:Deg_states}.}\label{fig:GS_energy}
\end{figure}

\hspace*{5mm} By minimizing the energy functional (Eq.~\ref{GS}) we identified GS configurations and calculated GS energy in a broad parameter space. In Fig.~\ref{fig:GS_energy} we show the GS energy surfaces in the $D/|J_1|-h/|J_1|$ parameter plane for various values of the biquadratic exchange parameter $J_2/|J_1|.$ Superimposed are the lines separating different phases, which are presented in Table~\ref{tab:Deg_states} with the lists of the corresponding degenerate states and their energies\footnote{The phases \circled{4} and \circled{8} are present only at $h/|J_1|=0$.}. The analytical expressions for these phase boundaries can be obtained by pairwise equating the GS energies. We can see that, depending on the parameters values, the system can be in up to eighth different phases. In the asymptotic limits of the parameters, for large negative values of the single-ion anisotropy $D/|J_1|$ the system is in the nonmagnetic state ${\bf S_{\mathrm{GS}}}=\{0,0,0\}$, while for large positive values of $D/|J_1|$ the nonmagnetic spin states are suppressed and we can expect the behavior similar to that of the spin-1/2 model. On the other hand, for large magnetic fields $h/|J_1|$ the system adopts the magnetic state ${\bf S_{\mathrm{GS}}}=\{1,1,1\}$ with all the spins fully aligned with the field direction. In the mentioned cases of the large negative $D/|J_1|$ (\circled{1}) and large $h/|J_1|$ (\circled{7}) the degeneracy is lifted completely. On the other hand, there is still three-fold (\circled{2},\circled{3},\circled{6}) and six-fold (\circled{4},\circled{5},\circled{8}) degeneracy if the parameters are of intermediate and relatively small values, respectively. We note that at the boundaries the degeneracy is further increased due to contributions of states from the neighboring phases and there are up to five points in which even three different phases coincide.

\begin{table}[t!]
\caption{Ground-state phases identified by Eq.~(\ref{GS}) with the corresponding degenerate states ${\bf S_{\mathrm{GS}}}$ and energies $f_{\mathrm{GS}}({\bf P})$, at the parameters values ${\bf P} = (J_2/|J_1|,D/|J_1|,h/|J_1|)$.}
\label{tab:Deg_states}
\centering
\begin{tabular}{c|c|c}
Phase & ${\bf S_{\mathrm{GS}}}=\{S_1,S_2,S_3\}$ & $f_{\mathrm{GS}}({\bf P})$ \\ \hline
\circled{1} & $\{0,0,0\}$ & $0$\\
\circled{2} & $\{0,0,1\};\{0,1,0\};\{1,0,0\}$ & $-\frac{D}{3|J_1|}-\frac{h}{3|J_1|}$ \\
\circled{3} & $\{0,1,1\};\{1,0,1\};\{1,1,0\}$ & $1-\frac{J_2}{|J_1|}-\frac{2D}{3|J_1|}-\frac{2h}{3|J_1|}$ \\
\circled{4} & $\{-1,0,0\};\{0,-1,0\};\{0,0,-1\};\{0,0,1\};\{0,1,0\};\{1,0,0\}$ & $-\frac{D}{3|J_1|}$ \\
\circled{5} & $\{-1,0,1\};\{-1,1,0\};\{0,-1,1\};\{0,1,-1\};\{1,-1,0\};\{1,0,-1\}$ & $-1-\frac{J_2}{|J_1|}-\frac{2D}{3|J_1|}$ \\
\circled{6} & $\{-1,1,1\};\{1,-1,1\};\{1,1,-1\}$ & $-1-\frac{3J_2}{|J_1|}-\frac{D}{|J_1|}-\frac{h}{3|J_1|}$ \\
\circled{7} & $\{1,1,1\}$ & $3-\frac{3J_2}{|J_1|}-\frac{D}{|J_1|}-\frac{h}{|J_1|}$ \\
\circled{8} & $\{-1,-1,1\};\{-1,1,-1\};\{-1,1,1\};\{1,-1,-1\};\{1,-1,1\};\{1,1,-1\}$ & $-1-\frac{3J_2}{|J_1|}-\frac{D}{|J_1|}$ \\
\end{tabular}
\end{table}

\begin{figure}[t]
\centering
		\subfigure{\includegraphics[scale=0.53]{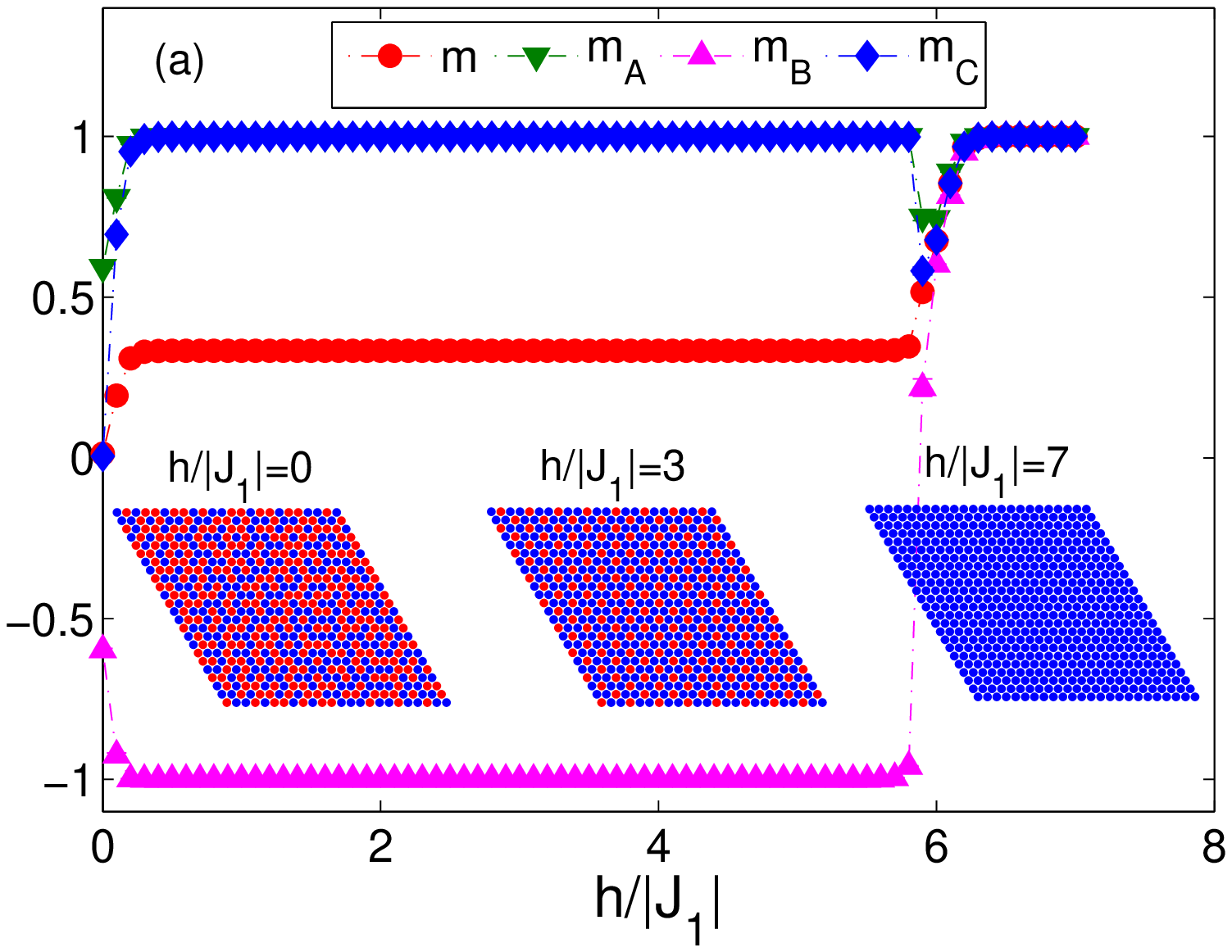}\label{fig:M-h_J2_0_D_1}}
    \subfigure{\includegraphics[scale=0.53]{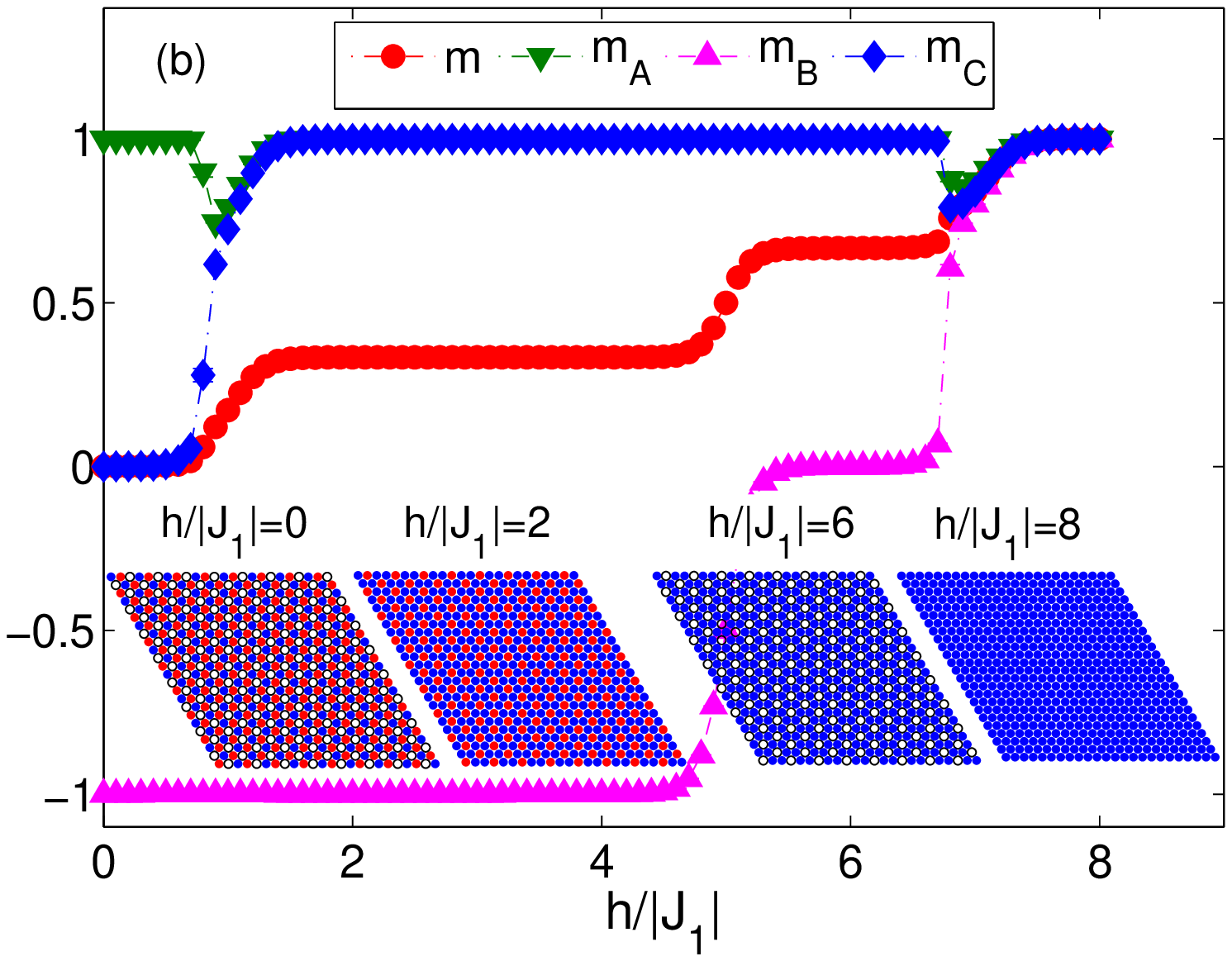}\label{fig:M-h_J2_0_D_-1}}
    \subfigure{\includegraphics[scale=0.53]{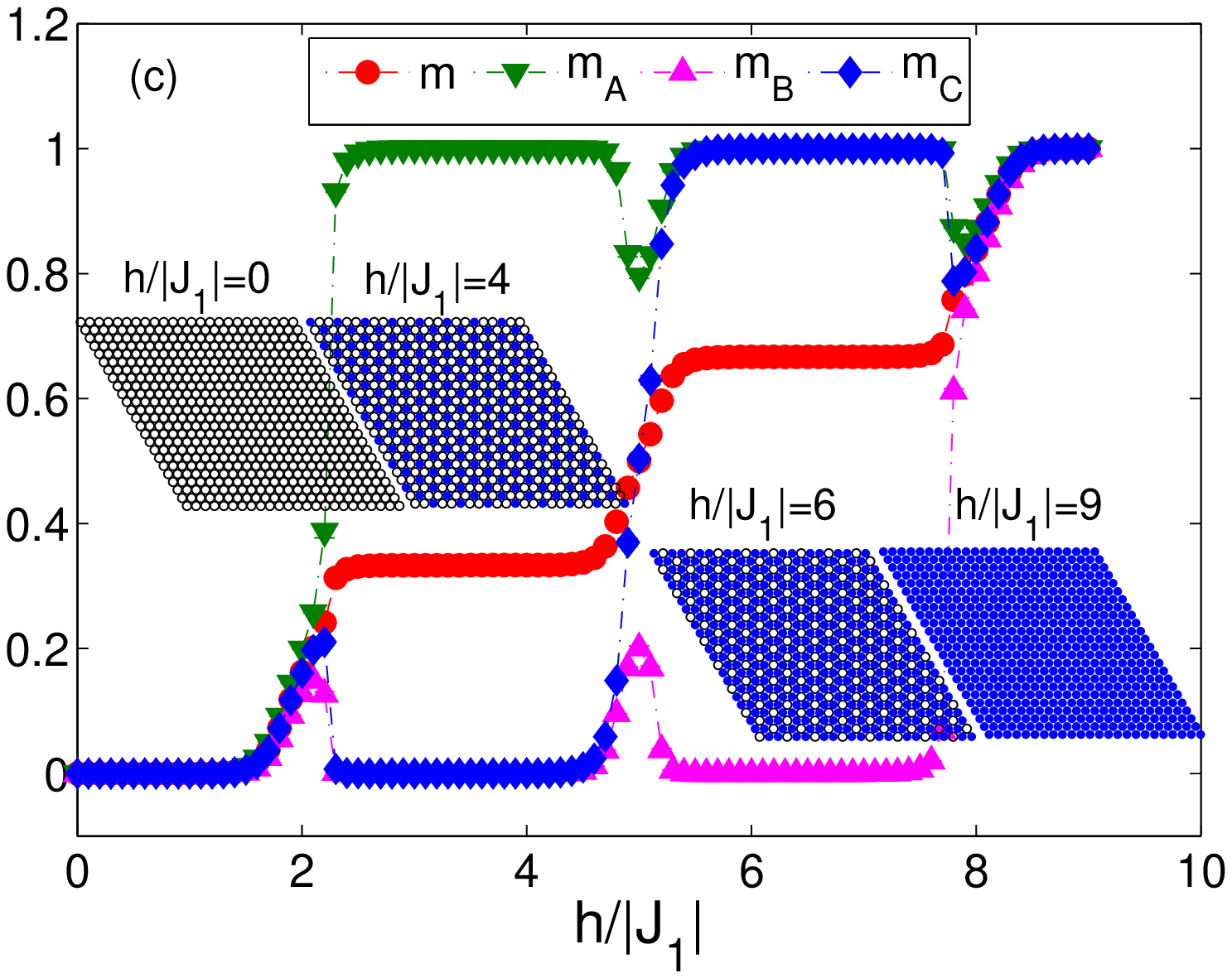}\label{fig:M-h_J2_0_D_-2}}
\caption{(Color online) Field-dependencies of the total magnetization $m$ and sublattice magnetizations $m_{\mathrm{A}}$, $m_{\mathrm{B}}$ and $m_{\mathrm{C}}$ at the temperature $t=0.1$ for $J_2/|J_1|=0$ and (a) $D/|J_1|=1$, (b) $D/|J_1|=-1$ and (c) $D/|J_1|=-2$. The snapshots in the respective figures illustrate typical spin configurations in different phases. The red (light gray), blue (dark gray) and white circles represent the spin states $-1$, $+1$ and $0$, respectively.}\label{fig:MC_mag_J20}
\end{figure}

\begin{figure}[t]
\centering
		\subfigure{\includegraphics[scale=0.53]{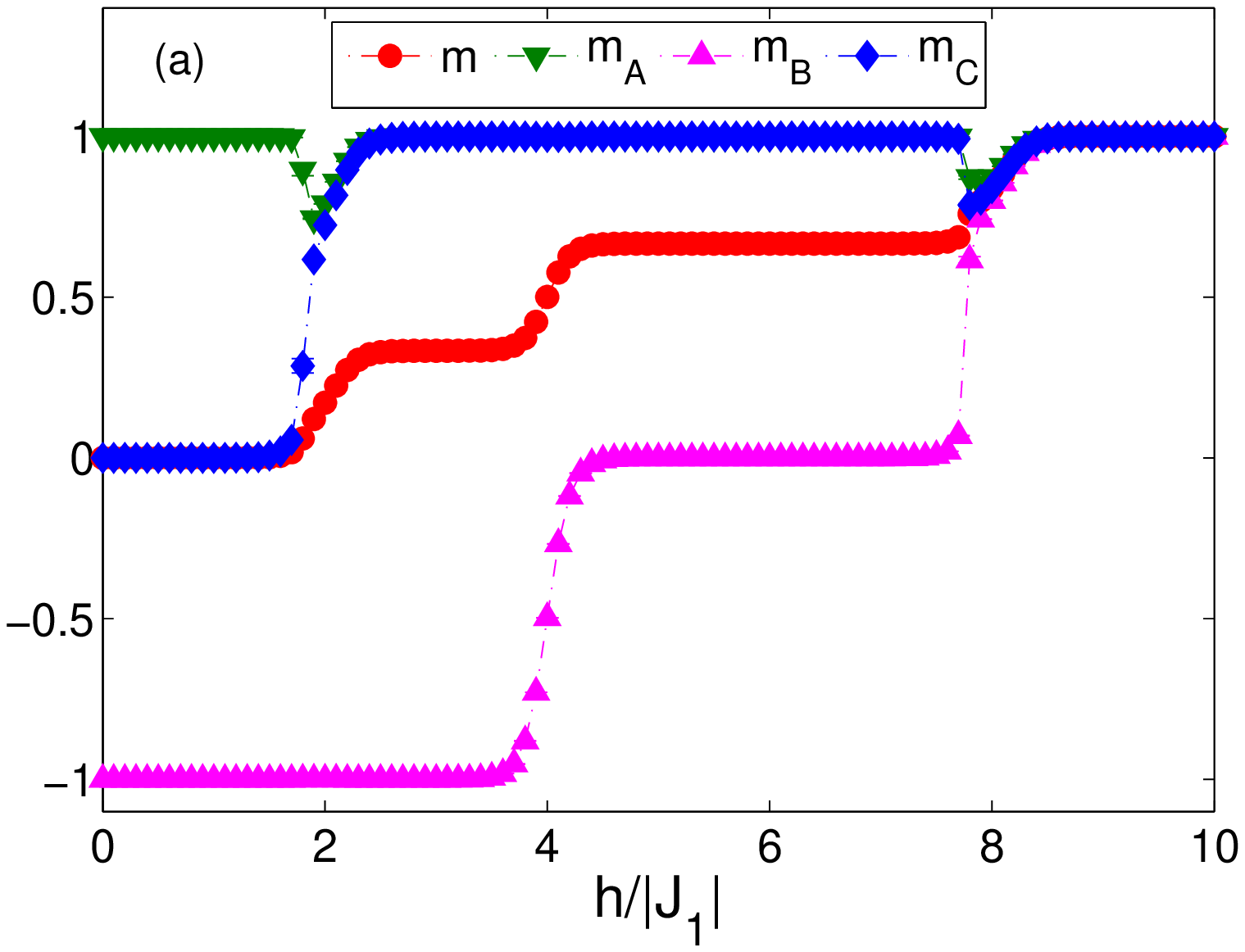}\label{fig:M-h_J2_-05_D_1}}
    \subfigure{\includegraphics[scale=0.53]{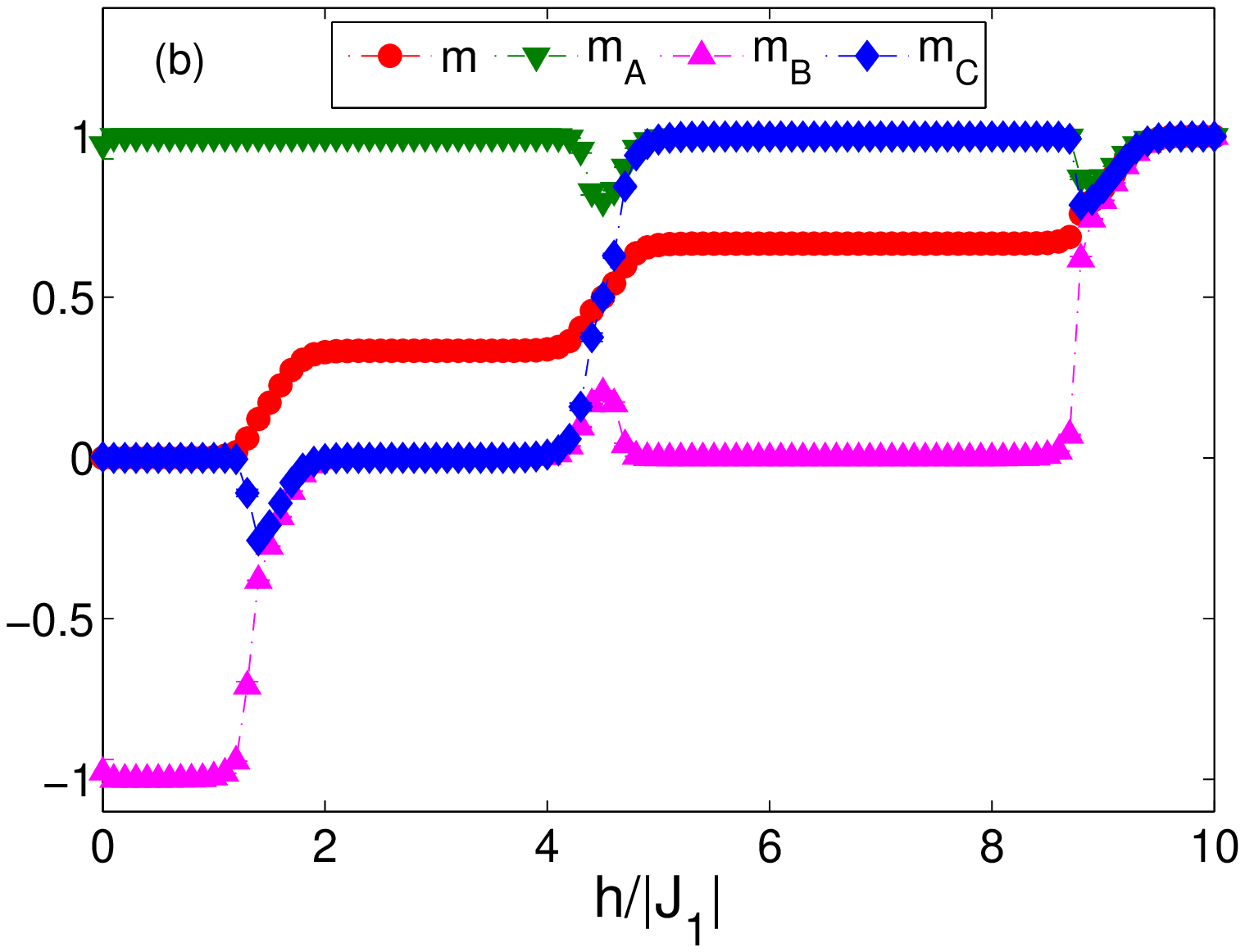}\label{fig:M-h_J2_-05_D_0}}
    \subfigure{\includegraphics[scale=0.53]{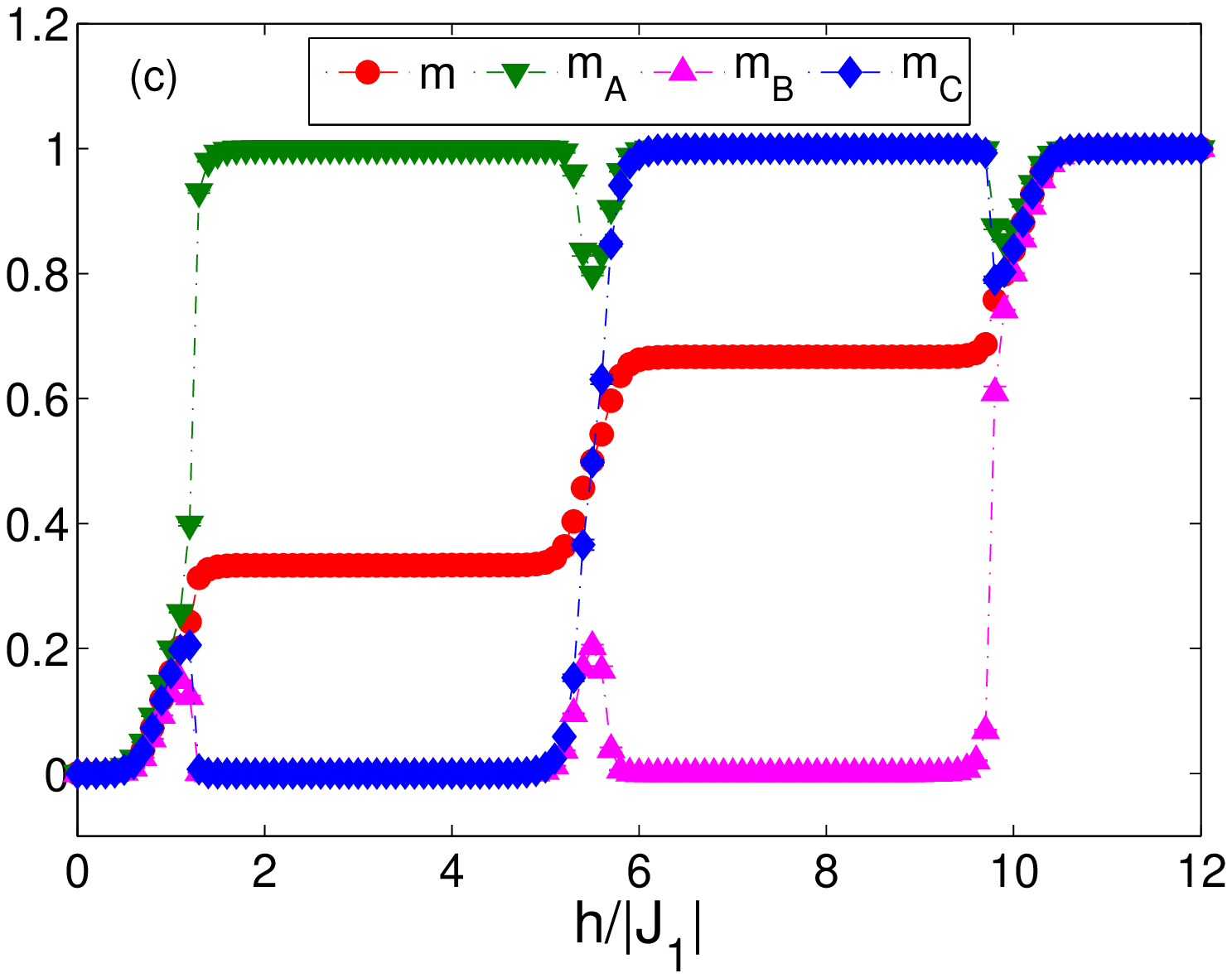}\label{fig:M-h_J2_-05_D_-1}}
\caption{(Color online) Field-dependencies of the total magnetization $m$ and sublattice magnetizations $m_{\mathrm{A}}$, $m_{\mathrm{B}}$ and $m_{\mathrm{C}}$ at the temperature $t=0.1$ for $J_2/|J_1|=-0.5$ and (a) $D/|J_1|=1$, (b) $D/|J_1|=0$ and (c) $D/|J_1|=-1$.}\label{fig:MC_mag_J2-05}
\end{figure}

\hspace*{5mm} With the respect to the above arguments on the behavior of the local spin patterns, we can expect the presence of interesting magnetic structures on the entire lattice in different regions of the parameter space. The character of the respective phases can be revealed by inspecting the  sublattice magnetizations $m_{\mathrm{A}}$, $m_{\mathrm{B}}$ and $m_{\mathrm{C}}$ as well as the total magnetization $m$, obtained from MC simulations. Their field-dependencies for a fixed temperature $t=0.1$ and the biquadratic exchange parameter $J_2/|J_1|=0$, and three different values of the single-ion anisotropy parameter are plotted in Fig.~\ref{fig:MC_mag_J20}. The inset spin snapshots in the respective figures illustrate typical spin configurations for different values of the field $h/|J_1|$, representing different phases. 
Positive values of the parameter $D$ tend to suppress nonmagnetic states and the spin system is expected to behave like a spin-1/2 TLIA. Namely, it is expected to show no long-range order in zero field, the ferrimagnetic state of the type $(m_{\mathrm{A}},m_{\mathrm{B}},m_{\mathrm{C}})=(1,-1,1)$ with two sublattice magnetizations aligned parallel and one antiparallel to the field direction (i.e., 1/3 magnetization plateau) within $h/|J_1| \in (0,6)$, and the ferromagnetic state with all three sublattice magnetizations aligned with the field direction for $h/|J_1| > 6$~\cite{metcalf,schick,netz,zuko}. This scenario is supported by the plots presented in Fig.~\ref{fig:M-h_J2_0_D_1} for $D/|J_1|=1$. On the other hand, negative values of the parameter $D$ tend to enhance nonmagnetic states. However, for relatively small magnitudes, more specifically for $J_2/|J_1|=0$ it is within $-3/2<D/|J_1|<0$ hence also for $D/|J_1|=-1$ shown in Fig.~\ref{fig:M-h_J2_0_D_-1}, in zero field the minimum energy is realized if nonmagnetic states are restricted to only one sublattice, while the remaining two sublattices are ordered antiferromagnetically, i.e., the structure of the type $(1,-1,0)$. As the field increases, the structure evolves first to the type $(1,-1,1)$, then to $(1,0,1)$ and finally for sufficiently large fields to $(1,1,1)$. Thus, the total magnetization $m$ features four plateaus with the values $m_1=0$, $m_2=1/3$, $m_3=2/3$ and $m_4=1$, corresponding to the respective magnetic structures. Finally, for negative $D$ of large magnitudes, namely for $J_2/|J_1|=0$ it is for $D/|J_1|<-3/2$ and thus also for $D/|J_1|=-2$ shown in Fig.~\ref{fig:M-h_J2_0_D_-2}, the nonmagnetic state $(0,0,0)$ is the ground state at small enough fields. The increasing field then brings the system successively to the states of the types $(1,0,0)$, $(1,0,1)$ and $(1,1,1)$. This again splits the magnetization curve $m$ into four plateaus of the same heights as for the case $D/|J_1|=-1$, albeit now the steps correspond to generally different magnetic structures. \\
\hspace*{5mm} In Fig.~\ref{fig:MC_mag_J2-05} we present similar plots also for a non-zero value of the biquadratic exchange interaction parameter, namely for $J_2/|J_1|=-0.5$, and three different values of the single-ion anisotropy parameter $D/|J|=-1,0,$ and $1$. The phases involved are the same as for the case of $J_2/|J_1|=0$ and, therefore, we do not illustrate their nature again by showing the snapshots. Nevertheless, the sequence of the phases through which the system passes as the field increases can be different, such as $\circled{5} \rightarrow \circled{2} \rightarrow \circled{3} \rightarrow \circled{7}$ for $D/|J|=0$, which cannot be observed at any value of $D/|J|$ if $J_2/|J_1|=0$.  

\section{Conclusions}
\hspace*{5mm} We have shown that the geometrically frustrated BEG model in an external magnetic field displays a relatively rich variety of different ground-state phases in the parameter space. They result from the large degeneracy, which is lifted in several stages with the increasing influence of the respective parameters. For selected parameter values we confirmed the existence of multiple magnetic structures, reflected in the step-wise field-dependence of the magnetization curves, in the Monte Carlo simulations at very low temperature. We believe that the present results will stimulate further study of the effects of thermal excitations on the individual ground-state spin arrangements and to establish entire finite-temperature phase diagrams. 

\section*{Acknowledgments}
This work was supported by the Scientific Grant Agency of Ministry of Education of Slovak Republic (Grant No. 1/0234/12). The authors acknowledge the financial support by the ERDF EU (European Union European Regional Development Fund) grant provided under the contract No. ITMS26220120005 (activity 3.2.).


\begin{thebibliography}{30}
\bibitem{wann} G.H. Wannier, Phys. Rev. 79 (1950) 357.
\bibitem{step} J. Stephenson, J. Math. Phys. 11 (1970) 413.
\bibitem{naga} O. Nagai, S. Miyashita, T. Horiguchi, Phys. Rev. B 47 (1993) 202.
\bibitem{yama} Y. Yamada, S. Miyashita, T. Horiguchi, M. Kang, O. Nagai, J. Magn. Magn. Mater. 140-144 (1995) 1749.
\bibitem{lipo} A. Lipowski, T. Horiguchi, D. Lipowska, Phys. Rev. Lett. 74 (1995) 3888.
\bibitem{metcalf} B.D. Metcalf, Phys. Lett. 45A (1973) 1.
\bibitem{schick} M. Schick, J.S. Walker, M. Wortis, Phys. Rev. B 16 (1977) 2205.
\bibitem{netz} R.R. Netz, A.N. Berker, Phys. Rev. Lett. 66 (1991) 377.
\bibitem{zuko} M. \v{Z}ukovi\v{c}, M. Borovsk\'{y}, A. Bob\'{a}k, Physics Letters A 374 (2010) 4260.
\bibitem{kaya} H. Kaya, A.N. Berker, Phys. Rev. E 62 (2000) R1469.
\bibitem{zuko1} M. \v{Z}ukovi\v{c}, M. Borovsk\'{y}, A. Bob\'{a}k, J. Magn. Magn. Mater. 324 (2012) 2687.
\bibitem{cape} H.W. Capel, Physica 32 (1966) 966.
\bibitem{blum} M. Blume, Phys. Rev. 141 (1966) 517.
\bibitem{beg} M. Blume, V.J. Emery, R.B. Griffiths, Phys. Rev. A 4 (1971) 1071.
\bibitem{berk1} A.N. Berker, M. Wortis, Phys. Rev. B 14 (1976) 4946.
\bibitem{chen} H.H. Chen, P.M. Levy, Phys. Rev. B 7 (1973) 4267.
\bibitem{netz1} R.R. Netz, A.N. Berker, Phys. Rev. B 47 (1993) 15019.
\bibitem{wang} Y.L. Wang, C. Wentworth, J. Appl. Phys. 61 (1987) 4411.
\bibitem{wang1} Y.L. Wang, F. Lee, J.D. Kimel, Phys. Rev. B 36 (1987) 8945.
\bibitem{host} W. Hoston, A.N. Berker, Phys. Rev. Lett. 67 (1991) 1027.
\bibitem{tuck} J.W. Tucker, T. Balcerzak, M. Gzik, A. Sukiennicki, J. Magn. Magn. Mater. 187 (1998) 381.
\bibitem{grig} G. Grigelionis, A. Rosengren, Physica A 208 (1994) 287.
\bibitem{akhe} A.Z. Akheyan, N.S. Ananikian, J. Phys. A 29 (1996) 721.
\bibitem{erdi} A. Erdinc, O. Canko, E. Albayrak, J. Magn. Magn. Mater. 303 (2006) 185.
\bibitem{maha} G.D. Mahan, S.M. Girvin, Phys. Rev. B 17 (1978) 4411.



\end{thebibliography}
\end{document}